\documentclass[aip,apl,reprint]{revtex4-1}
\usepackage{graphicx}
\usepackage{amsmath}
\usepackage{dsfont}
\usepackage{enumerate}
\newcommand{\ve}{\varepsilon}
\begin{document}
\title{Snake states and Klein tunneling in a graphene Hall bar with a pn-junction}
%
\author{M. Barbier}
\affiliation{Departement Fysica, Universiteit Antwerpen, Groenenborgerlaan 171, B-2020 Antwerpen, Belgium\\}%
\author{G. Papp}
\affiliation{Institute of Physics and Electrotechnics, University of West Hungary, Bajcsy-Zsilinszky {\'u}t 4-6, H-9400 Sopron, Hungary}
\affiliation{Department of Theoretical Physics, University of Szeged, Tisza Lajos k{\"o}r{\'u}t. 84-86, H-6720 Szeged, Hungary}
\author{F. M. Peeters}
\affiliation{Departement Fysica, Universiteit Antwerpen, Groenenborgerlaan 171, B-2020 Antwerpen, Belgium\\}%
\begin{abstract}
The Hall ($R_H$) and bend ($R_B$) resistance of a graphene Hall bar structure 
containing a pn-junction are calculated when in the ballistic regime. 
The simulations are done 
using the billiard model. 
Introducing a pn-junction--dividing the Hall bar geometry in two regions--leads 
to two distinct regimes exhibiting very different physics: 1) both regions are 
of n-type and 2) one region is n-type and the other p-type.
In regime (1) a `Hall plateau'--an enhancement of the resistance--appears for $R_H$. 
On the other hand, in regime (2), we found a negative $R_H$, which approaches zero for large B.
The bend resistance is highly asymmetric in regime (2) and the resistance increases with 
increasing magnetic field B in one direction while it reduces to zero in the other direction.
\end{abstract}
%
\maketitle
%
%
In graphene, a single flat layer of hexagonal carbon atoms, the carriers act as a 2D ultra-relativistic fermion gas.
Among interesting differences with standard two-dimensional electron gases (2DEGs) we should mention: the extra-ordinary Quantum Hall effect, Klein paradox \cite{Fuchs}, and Zitterbewegung\cite{Rusin_review},
as some of the most striking characteristics of this new material.
Although for most of these effects only indirect signatures were found in experiments up to now, graphene's behaviour as a relativistic analogue has far-reaching consequences in electronic devices when based on this material.
In this context, the pn-junction is an interesting system 
because it exhibits 
Klein tunneling as well as the analogue of a negative refraction index.
Moreover the phenomen of `snake states' along the pn-interface
has been predicted \cite{Davies, Carmier}.
Actually at such a pn-junction one has a combination of `skipping orbits' when electrons
arriving at the pn-junction are reflected and `snake states' when they exhibit negative refractive index behaviour.
Along with theoretical proposals \cite{Davies}, experiments on such systems 
were undertaken recently \cite{Williams, Lohmann}.
This motivated us to investigate the response of a graphene Hall bar \cite{Weingart, Mayorov} containing a pn-junction.
In previous work on mesoscopic Hall bars Beenakker \emph{et al.} \cite{Beenakker_billiard} 
proposed a classical model to describe the magnetic response.
Such Hall probes have been applied as noninvasive probes for local magnetic fields \cite{Novolesov_Qdot_hallbar,Li}
to investigate e.g. mesoscopic superconducting disks and ferromagnetic particles.
In this work we investigate a Hall-bar made of graphene with a pn-junction along one of 
its axes in the ballistic regime using the billiard model.

%
\begin{figure}[htbp]
	\includegraphics{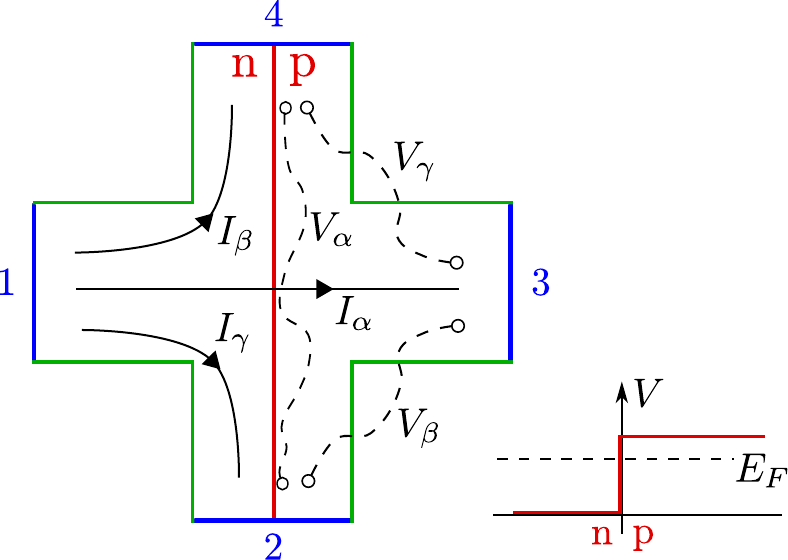}
  \caption{(Color online) A sketch of the Hall-bar geometry, where the pn-junction is represented by the red (dark grey) line, and the contacts are shown in blue (black) having a chemical potential $\mu_i$. Three configurations are possible for the voltage and current probes: $\alpha$ (Hall resistance measurement), and $\beta$ and $\gamma$ (bend resistance measurements). The inset shows the potential profile.
}\label{fig:1}
\end{figure}
The system we investigate, shown in Fig.~\ref{fig:1}, is a Hall bar with four 
identical leads, with in the middle a pn-junction dividing the Hall cross in an 
n-doped region and a p-doped region. 
The doping can be realized electrically by applying gate potentials $V_p$ and $V_n$ 
on the p- and n-region respectively.
We modelled this by a shift in energy $E_F \rightarrow E_F - V_n$ and define 
$V = V_p - V_n$, which is equivalent to have $V_n = 0$.
We use the following dimensionless units throughout the work, $R^* = R/R_0$, 
with $R_0 = \frac{h}{4 e^2} \frac{\hbar v_F}{ |E_F| W}$ \cite{note:2}, 
$B^* = B/B_0$, with $B_0 = \frac{|E_F|}{e v_F W}$,	$r^*_c = r_c/W = (1-\xi)/B^*$, with $\xi = V/E_F$, 
where $W$ is the width of the channels, $v_F \approx 10^6$m/s is the Fermi-velocity, $e$ is the charge of the electron.
Our numerical simulations are done for 
typical values of $W = 1\mu$m and $E_F = 50$meV, which results in 
$R_0 = 6.08\times 10^2 \frac{h}{4 e^2}$ and $B_0 = 0.05$T.
%
In order for our approach to be valid, we must require that: (1) we are in the 
ballistic regime, meaning that the mean free path $l_e$ is larger than the width of the 
channels $W$, and (2) the system can be described classically, implying that, a) 
the confinement quantization $E_q = \hbar v_F/W $ and b) the Landau level quantization $\hbar \omega_D = \sqrt{2} \hbar v_F/l_B$ with the magnetic length $l_B = \sqrt{\hbar/e B}$ are small compared to the thermal fluctuations $k_B T$.
Since in graphene the temperature dependence of $l_e$ is found to be small within a broad 
range of temperature, and $l_e$ is of order of microns for clean graphene samples, 
see e.g. Ref.~\cite{Mayorov},
 (1) is satisfied for not too low temperatures, and for the used parameters.
%

We calculate the Hall and bend resistances according to the Landauer-B\"{u}ttiker 
theory \cite{Buttiker} where the resistances are defined as
\begin{equation}\label{eq:res}
	R_{mn,kl} = \frac{V_{kl}}{I_{mn}} = R_0\frac{T_{km} T_{ln} - T_{kn} T_{lm}}{D},
\end{equation}
with the voltage $V_{kl} = (\mu_k - \mu_l)/e$ measured between terminals ($k$, $l$) and 
a current $I_{mn}$ is flowing between ($m$, $n$). $T_{ij}$ are the transmission matrix elements giving the probability 
for an electron injected from terminal $j$ to end up in terminal $i$, 
and $D$ is a (sub)determinant of the transmission matrix $T$ which is symmetric 
in the magnetic flux: $D(B) = D(-B)$).

One has the following exact symmetries: (1)~$R_{mn,kl} = -R_{mn,lk} = -R_{nm,kl}$, and (2)~the reciprocity relation $R_{mn,kl}(B) = R_{kl,mn}(-B)$.
There are three resistances of interest: the Hall resistance $R_H = R_\alpha = R_{13,24}$, the bend resistances 
$R_B = R_\beta = R_{14,32}$ and $R_\gamma = R_{12,43}$, and their counterparts (denoted by a prime) obtained by 
switching the voltage and current probes, where the latter are related to the 
former by the relation $R_{i}(-B) = -R'_{i}(B)$.
These resistances are defined by 
$R_{i} = V_{i}/I_{i}$ $(i = \alpha, \beta, \gamma)$,
as shown in Fig.~\ref{fig:1}.

To obtain the transmission matrix elements we use the semi-classical \emph{billiard model} \cite{Beenakker_billiard}.
$T_{ij}$ are found by injecting
a large number $N$ of electrons from terminals $j$ and tracking down which ratio ends up in terminals $i$. 
The probability to be injected from a terminal $j$ is according to Ref.~\cite{Beenakker_billiard} uniformly distributed over the channel width, and has an angular distribution $P(\alpha)=\frac{1}{2}\cos(\alpha)$. 
In graphene it also depends on the DOS which changes with the parameter $\xi = E_F/V$. 
Because in the system under consideration the DOS is not the same in every channel (and even not homogeneous inside channels 2 and 4) we incorporate this by a factor $\ve_j(x) = |1-\xi_j(x)|$ for injection from position $x$ at terminal $j$.

The trajectories of the electrons are assumed to be ballistic and the electron moves along a circular orbit with cyclotron radius $r_c$.
Unlike in a standard 2DEG where $r^{classical}_c = \frac{p}{eB} = \frac{m v^2}{eBv}$,
in graphene we have a linear dispersion relation $E_{kin} = v_F p$ and we obtain instead
\begin{equation}\label{eq:Rc}
	r_c = \frac{p_F}{e|B|} = \frac{E_{kin}}{e v_F |B|} = \frac{|E_F-V|}{e v_F |B|},
\end{equation}
a cyclotron radius which can be tuned by changing the magnetic 
field strength $B$ or the potential $V$. 
To investigate the behavior of an electron incident on the pn-junction, 
we calculate the transmission and reflection probabilities on the pn-junction quantum 
mechanically (according to the Dirac model), as well as, how the particle is 
deflected by the junction. 
The direction of the particle will change on transmission through the junction. 
If the incident angle is $\phi_i$ then in case of reflection, the reflected angle is $\phi_r=\phi_i$. 
For transmission the angle $\phi_t$ of the 
transmitted electron is related to the incident one as
\begin{equation}
	p_t \sin(\phi_t) = p_i \sin(\phi_i),
\end{equation}
where $p_t$ and $p_i$ are the corresponding momenta in the respective regions, 
if the region has potential V then $p=(E-V)/v_F$. 
In the case $|\sin(\phi_t)|>1$, total reflection occurs and the reflection 
probability $R=1$. Otherwise $R$ is given by\cite{Davies}
\begin{equation}
	R=\left| \sin\left(\frac{\phi_1+\phi_2}{2}\right) / \cos\left(\frac{\phi_1-\phi_2}{2}\right)\right|^2.
\end{equation}
Thus when an electron hits the pn-junction, it has a probability to be reflected
or to be transmitted.
%

Although the four-fold symmetry is lost in our system, we still have the following symmetries:
\begin{enumerate}[1)]\label{eq:Tsym}
	\item{$T_{ij}(B) = T_{ji}(-B)$, the Onsager relation},
	\item{$T_{j2}(B) = T_{j4}(-B)$, with $j = 1,3$},
	\item{$T_{13}(B) = T_{13}(-B)$},
\end{enumerate}
where the last two are specifically for our pn-junction set-up.
Therefore, both the Hall and bend resistances obey
\begin{enumerate}[1)]\label{eq:Rsym2}
	\item{$R_\gamma(-B) = -R_\beta(B)$},
	\item{$R_\alpha(-B) = -R_\alpha(B)$},
\end{enumerate}
Hence it suffices for the Hall resistance to only plot $R_\alpha(B)$ for positive values of the magnetic field.
Moreover for the bend resistance we choose to plot $R_\beta(B)$.
%

\begin{figure}[htbp]
	\includegraphics{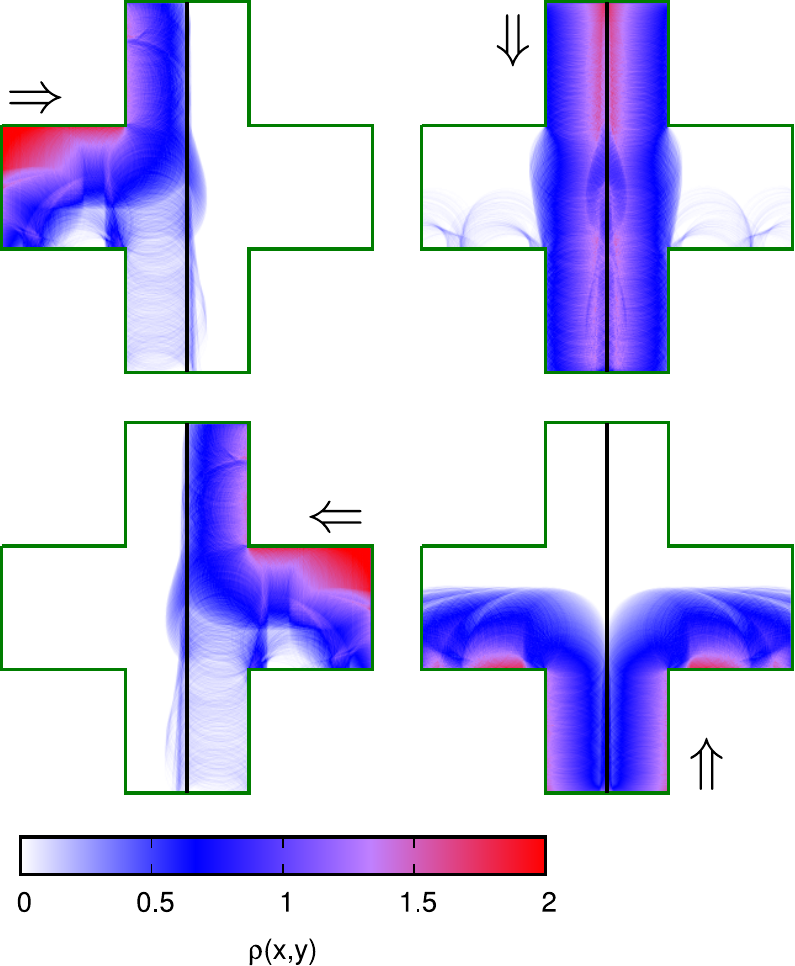}
  \caption{(Color online) The electron density $\rho(x,y)$ inside the Hall cross for $B^* = 3$, $V/E_F = 2$ for
carrier injection from the four different leads as indicated by the arrows.
	}\label{fig:2}
\end{figure}

To understand the influence of the pn-junction on the 
resistances we plot the electron density $\rho(x,y)$ in
Fig.~\ref{fig:2} for $B^* = 3$, $V/E_F = 2$. 
For this $B$ the cyclotron radius $r_c = W/3$ is 
equal in both regions, but with opposite direction of motion for the electrons. 
Notice that $r_c = W/3$ 
is large enough that the charge carriers will reach the pn-junction, 
but not the other edge of the channel without 
scattering (remember the width of the channels is $W$). 
There is symmetry for injection from lead 1 and 3 because of equal cyclotron 
radius in both regions.
For injection in the first lead we see that due to the
negative refraction index there is a nonzero $T_{21}$.
Comparing injection from lead 2 and 4 we notice that the
snake states are only travelling from 4 to 2 and not in the
opposite direction.

\begin{figure}[htbp]
	\includegraphics{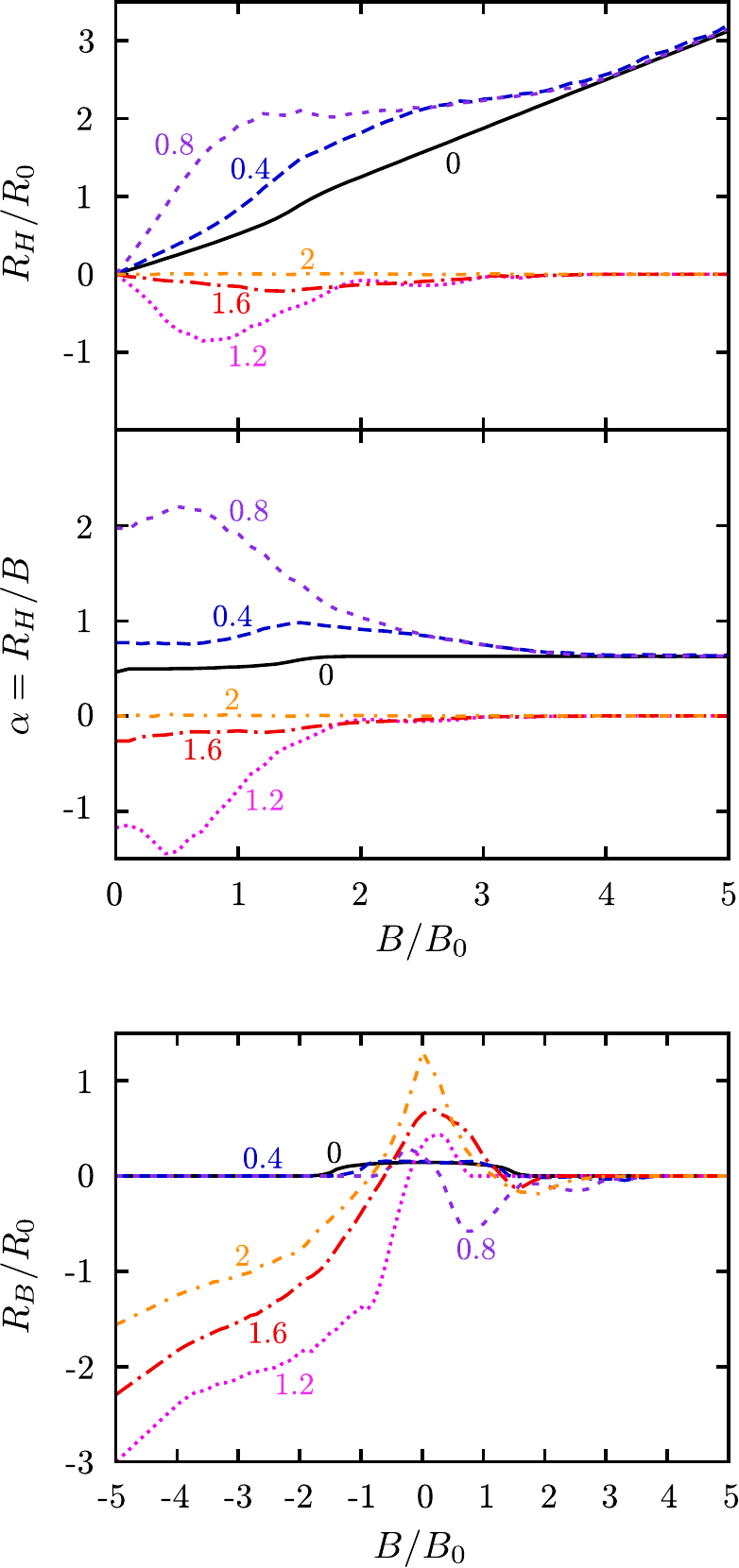}
  \caption{(Color online) (a) Hall resistance, (b) Hall coefficient, and (c) the bend resistance, 
versus the magnetic field strength are shown for various values of 
$\xi = V/E_F = 0; 0.4; 0.8; 1.2; 1.6; 2$, given by the labels next to the curves.
	}\label{fig:3}
\end{figure}

In Fig.~\ref{fig:3} the (a) Hall resistance, (b) Hall coefficient, and (c) 
the bend resistance are shown for various values of $\xi = V/E_F$. 
Two cases are clearly distinct: (1) $0 < \xi < 1$, such that the potential barrier is lower than the Fermi-energy and both regions are of n-type, and (2) $\xi > 1$ where the right region is p-type, which reverts the circling direction of the electrons and makes the refraction index negative allowing for Klein tunneling in graphene.
From this figure we see that the Hall resistance shows for small $0 < \xi < 1$,
a plateau, while for $\xi > 1$ it is negative and approaches zero with increasing $\xi$.
The plateau in $R_H$ is due to the pn-junction which guides the electrons making 
it difficult for them to cross the junction \cite{note:1}.
If $\xi$ is increased above $1$, the electrons in the p region will turn in the opposit direction,
therefore they can bend towards a different channel and the pn-junction 
will collimate the electrons\cite{Cheianov} diminishing the Hall resistance 
with increasing $\xi = V/E_F$ until it vanishes at $V = 2 E_F$ where $r_c$ is the same in both regions.
From the Hall coefficient we can see that unlike the case without pn-junction, 
the Hall coefficient is negative and decreases for high magnetic fields in the 
case $\xi>1$, while for $\xi<1$ it approaches 
asymptotically the value for $\xi=0$ with increasing $B$.
The bend resistance on the contrary becomes highly asymmetric, with increasing 
$\xi = V/E_F$ and is strongly enhanced by the presence of the pn-junction. 
In particular, notice that in regime (2) $|R_B|$ increases for large and negative $B$.
To understand this we observe from Eq.~(\ref{eq:res}) that $R_B \propto T_{31}T_{24} - T_{34}T_{21}$.
Notice further that for large and negative $B$, $T_{31} \approx 0$ 
while $T_{21}$ and $T_{34}$ are nonzero because the right region is of p-type.

%
In summary, we investigated the Hall and bend resistance of a Hall cross 
made of graphene containing a pn-junction.
The Hall resistance exhibits a `last Hall plateau', 
as in a standard 2DEG when the Hall bar has rounded corners, in the case of $\xi < 1$
where the junction serves as a guiding center for the electrons.
For $\xi > 1$, the Hall resistance is qualitatively very different:
i) its sign is opposite, ii) no Hall plateau is present but rather a local minimum,
iii) $|R_H|$ diminishes both with increasing $\xi$ and $B$, and iv) for $\xi = 2$, $R_H = 0$.
The bend resistance is highly asymmetric in $B$ for large $V$.
\acknowledgments
This work was supported by IMEC, 
the Flemish Science Foundation (FWO-Vl), 
the Belgian Science Policy (IAP), 
and the ESF-EuroGRAPHENE project CONGRAN.

\begin{thebibliography}{99}
\bibitem{note:2} {Normally $R_0 = \frac{h}{2e^2}$ for a standard 2DEG, but since graphene has a non constant density of states (DOS) equal to $\rho(E) = \frac{4|E|}{2\pi(\hbar v_F)^2}$, we have $R_0 \propto 1/E_F$.}
\bibitem{note:1} This is very similar to the guiding found in Ref.~\cite{Beenakker_billiard}, where the guiding 
occurs due to the rounding of the edges of the Hall cross.
\end{thebibliography}
\end{document}